\documentclass[twocolumn,superscriptaddress,aps,floatfix]{revtex4}
\usepackage{amsmath}
\usepackage{amssymb}
\usepackage{graphicx}
\usepackage{color}
\usepackage[colorlinks=true,linkcolor=blue,citecolor=blue]{hyperref}
\begin{document}

\title{Evolution of Magnetic Interactions in Sb-substituted MnBi$_{2}$Te$_{4}$}

\author{S. X. M. Riberolles}
\affiliation{Ames Laboratory, Ames, IA, 50011, USA}

\author{Q. Zhang}
\affiliation{Oak Ridge National Laboratory, Oak Ridge, TN, 37831, USA}

\author{Elijah Gordon}
\affiliation{Ames Laboratory, Ames, IA, 50011, USA}

\author{N. P. Butch}
\affiliation{NIST Center for Neutron Research, National Institute of Standards and Technology, Gaithersburg, MD 20899, USA}

\author{Liqin Ke}
\affiliation{Ames Laboratory, Ames, IA, 50011, USA}

\author{J.-Q.~Yan}
\affiliation{Oak Ridge National Laboratory, Oak Ridge, TN, 37831, USA}

\author{R.~J.~McQueeney}
\affiliation{Ames Laboratory, Ames, IA, 50011, USA}
\affiliation{Department of Physics and Astronomy, Iowa State University, Ames, IA, 50011, USA}

 \date{\today}

\begin{abstract}
The Mn(Bi$_{1-x}$Sb$_{x}$)$_{2}$Te$_{4}$ series is purported to span from antiferromagnetic (AF) topological insulator at $x=0$ to a trivial AF insulator at $x=1$.  Here we report on neutron diffraction and inelastic neutron scattering studies of the magnetic interactions across this series. All compounds measured possess ferromagnetic (FM) triangular layers and we find a crossover from AF to FM interlayer coupling near $x=1$ for our samples.  The large spin gap at $x=0$ closes rapidly and the average FM exchange interactions within the triangular layer increase with Sb substitution.  Similar to a previous study of MnBi$_2$Te$_4$, we find severe spectral broadening which increases dramatically across the compositional series.  In addition to broadening, we observe an additional sharp magnetic excitation in MnSb$_{2}$Te$_{4}$ that may indicate the development of local magnetic modes based on recent reports of antisite disorder between Mn and Sb sublattices.  The results suggest that both substitutional and antisite disorder contribute substantially to the magnetism in Mn(Bi$_{1-x}$Sb$_{x}$)$_{2}$Te$_{4}$.
 \end{abstract}

\maketitle

\section{Introduction}
Antiferromagnetic topological insulators (AFTI) are predicted to provide a platform for novel topological phases, such as quantum anomalous Hall (QAH) insulators, axion insulators, or Weyl semimetals \cite{Moore10}. Layered magnetic materials based on the MnBi$_{2}$Te$_{4}$ (MBT) prototype, which consist of stacks of Bi-Te layers containing inverted topological bands and magnetic Mn layers, has been predicted to belong to this class of materials \cite{Otrokov17,Zhang19,Otrokov18_2,Lee19,Yan19}.  The antiferromagnetic (AF) order of Mn moments consists of FM triangular layers with AF interlayer coupling, referred to as $A$-type AF order, with moments pointing perpendicular to the layers \cite{Yan19}. In principle, $A$-type magnetic structure provides access to novel topological phases.  For example, bulk materials and thin films with an even number of Mn layers are predicted to be axion insulators due to the preservation of product time-reversal and half-lattice translation.  On the other hand, QAH states are expected when time-reversal symmetry is broken in field-polarized thick films or thin films with an odd number of magnetic layers \cite{Otrokov19,Gong19,Zhang19} and have recently been observed \cite{Deng19}.  The search for QAH states is enabled by relatively weak-field metamagnetic transitions which allow access to canted, spin-flopped, or fully polarized magnetic structures \cite{Otrokov18_2,Lee19,Yan19}.

However, MnBi$_{2}$Te$_{4}$ presents some materials difficulties that may prevent its utilization as a platform for novel topological functionality.  The first issue is that the bulk electronic structure is not insulating, but rather hosts a high concentration of $n$-type charge carriers.  The second issue is that the expected gapping of topological electronic surface states has not been observed by angle-resolved photoemission spectroscopy\cite{Chen19, Hao19, HangLi19, Swatek20, Nevola20}, suggesting that the FM structure of Mn layers close to the surface of the sample is compromised by disorder or some other magnetic instability.  On the first issue, it has been recognized that chemical substitution of Sb for Bi, Mn(Bi$_{1-x}$Sb$_{x}$)$_{2}$Te$_{4}$ changes the carrier type from $n$-type to $p$-type and compositions with $x \approx 0.3$ have a relatively low carrier concentration ($n=3\times$10$^{18}$ cm$^{-3}$) suggesting that the chemical potential is in the gap \cite{Chen19_MST, Yan19}.  For full Sb substitution, the carrier concentration for MnSb$_{2}$Te$_{4}$ (MST) is hole-like with $n=4\times$10$^{20}$ cm$^{-3}$.  Ultimately, heavy Sb substitution decreases the average spin-orbit coupling (SOC) and density functional theory (DFT) calculations predict a closing of the inverted band gap.   This has been predicted to close at $x\approx 0.6$  resulting in a crossover from topological to normal insulator \cite{Chen19_MST}, although recent reports suggest that MST itself is a topological insulator \cite{Wimmer20}.  Thus, from perspective of the bulk electronic structure, the Mn(Bi$_{1-x}$Sb$_{x}$)$_{2}$Te$_{4}$ series can tailor both the chemical potential, the band gap and possibly the band inversion.

On the second issue, it has been observed that Sb substitution causes several changes in the magnetism whose ultimate effect on bulk or surface electronic structures are unclear.  Sb substitution results in a decrease of the saturated magnetization \cite{Yan19_2} and ordered magnetic moment obtained from neutron diffraction \cite{Murakami19}.  A picture is developing where this moment suppression may arise from AF coupling between inequivalent Mn ions caused by chemical antisite disorder between Mn and Sb sites.  Also, Sb substitution results in a suppression of spin-flop transition, indicating that the single-ion anisotropy or the interlayer exchange coupling (or both) are reduced \cite{Yan19_2}.  These reduced energy scales lower the energy barrier between FM and $A$-type AF ground states and, in fact, AF and FM ground states are both observed in MST dependent on growth conditions that may introduce vacancies, site disorder or strain \cite{Liu20, Wimmer20}.  The MST sample studied here was found to have a FM ground state.

To get a better understanding of the evolution of the magnetism with Sb substitution, we use neutron diffraction and inelastic neutron scattering (INS) to study the evolution of the magnetism in polycrystalline samples of Mn(Bi$_{1-x}$Sb$_{x}$)$_{2}$Te$_{4}$.  For our samples, we obtain three main conclusions from diffraction data: (1) heavy chemical disorder is present in Sb-rich samples in the form of antisite defects between Mn and Sb, (2) magnetism evolves from $A$-type AF to FM, consistent with a sign-change of the small  exchange coupling between septuple blocks, (3) coupling between Mn layers and Mn in the Bi/Sb layers within a septuple block is AF, which results in partial compensation of the ordered moment.  Inelastic scattering results also reveal several features that evolve with composition. We find that the spin gap of ~0.6 meV in MBT rapidly closes with Sb substitution, consistent with the observed decrease in the spin-flop field. We also find that the overall energy scale of the spin fluctuations increases substantially and spectral features at high energy have extreme lifetime broadening.  Such broadening may arise from the observed chemical disorder or the strong coupling of the magnetism and charge carriers.  Finally, we observe spectral features consistent with the development of a ferrimagnetic block layer in MST due to antisite mixing of Mn and Sb. In this case, we show that localized magnetic excitations will arise from strong AF coupling of Mn layers to antisite Mn in Sb layers. A comparison of these results to DFT calculations highlights that the magnitude and sign of the intra-septuple block and inter-septuple block magnetic interactions are very sensitive to the chemical and magnetic defect configurations.
 
 \section{Experimental Details}
 Powder samples of nominal Mn(Bi$_{1-x}$Sb$_{x}$)$_{2}$Te$_{4}$ with $x=$ 0, 0.25, 0.5, 0.75, and 1 were synthesized by annealing at 585$^{o}$C for a week the homogeneous stoichiometric mixture of the elements after quenching from 900$^{o}$C.  Neutron powder diffraction measurements were performed on the time-of-flight (TOF) powder diffractometer POWGEN at the Spallation Neutron Source, SNS Oak Ridge National Laboratory. Each sample had a mass of approximately 2.7 g and was loaded into standard cylindrical vanadium cans.  Data at 100 K were collected for a total proton charge of 5 coulombs. Inelastic neutron scattering experiments were performed on the Disk Chopper Spectrometer (DCS) at the NIST Center for Neutron Scattering.    Approximately six grams of powder sample for each composition were loaded into 1/2" diameter aluminum can and attached to the cold stick of a closed-cycle refrigerator for measurements below and above $T_{N}$ at $T=$ 4 K and 30 K, respectively, using incident neutron energies of $E_{i}=$ 3.55 and 9.09 meV.  Both the elastic and inelastic data from DCS were analyzed.  For the inelastic data, the intensities are plotted as $S(Q,E)/(1+n(E))$ where $Q$ is the momentum transfer, $E$ is the energy transfer,  and $n(E)=[\mathrm{exp}(E/k_{B}T)-1]^{-1}$ is the Bose population factor. This intensity is proportional to the imaginary part of the dynamical susceptibility times the square of the magnetic form factor, $f^{2}(Q)\chi''(Q,E)$.

\section{Analysis of diffraction data}
Neutron diffraction data taken above the ordering transition at $T=$ 100 K were used to determine the crystallinity and verify the doping concentration. Rietveld refinement techniques  employed with the GSAS-II software \cite{GSAS} confirm the $R\bar{3}c$ rhombohedral structure.  Additional diffraction peaks were observed to originate from MnTe and Bi$_2$Te$_3$ impurity phases at the level of 5--10\% volume fraction.  Overall, the refinements establish that the actual Sb concentration is in excellent agreement with the nominal value.  Refinements also find evidence for significant antisite mixing between Mn ($3a$) and Bi/Sb $6c$-sites, as reported previously \cite{Murakami19}.  Generally, the concentration of Mn at the $6c$ site increases with $x$ indicating that antisite becomes more likely with increased Sb substitution.  The concentration of antisite mixing reaches a maximum of ~16\% for the FM MST sample.  Refinements also find additional vacancies in the Mn $3a$ layer at a concentration of 5-8\% that does not seem to vary strongly with Sb substitution level. 

The site specific chemical formula can be written as (Mn$_{1-2y-z}$X$_{2y+z}$)(X$_{1-y}$Mn$_{y}$)$_2$Te$_4$. Here X$=$ Bi$_{1-x}$Sb$_x$ represents an average Bi/Sb ion with Sb concentration $x$.  Antisite mixing between $3a$ and $6c$ is represented by the concentration of $y$ Mn ions in a single X ($6c$) layer (and a corresponding concentration of $2y$ X ions in the Mn layer).  Finally, $z$ represents the concentration of additional X ions in the $3a$ Mn layer, thereby creating additional magnetic vacancies in the $3a$ layer beyond those created by antisite exchange.  The total magnetic dilution of the main Mn $3a$ layer is $2y+z$.  Table \ref{Rietveld} summarizes the chemical composition and configuration and Table \ref{Lattice} reports the lattice constants and atomic positions according to Rietveld refinement of the 100 K POWGEN data. 

\begin{table}
\caption {Chemical configuration of nominal Mn(Bi$_{1-x}$Sb$_{x}$)$_2$Te$_4$ samples as determined from Rietveld refinement of POWGEN neutron diffraction data at 100 K.  The chemical configuration is reported in terms of the relative concentration of Sb ($x$), antisite mixing concentration ($y$), additional vacancy concentration ($z$), and the total magnetic dilution of the Mn layer ($2y+z$).  This is encapsulated in either the site specific chemical formula (Mn$_{1-2y-z}$X$_{2y+z}$)(X$_{1-y}$Mn$_{y}$)$_2$Te$_4$ (where X$=$ Bi$_{1-x}$Sb$_x$ represents an average Bi/Sb ion) or the formula unit Mn$_{1-z}$(Bi$_{1-x}$Sb$_{x}$)$_{2+z}$Te$_4$.}
\renewcommand\arraystretch{1.25}
\centering
\begin{tabular}{ c | c | c | c | c | c}
\hline\hline
~ $x_{\rm nom}$ ~ & ~$x$~ & ~ $y$ ~ & ~$z$ ~ & $2y+z$ & Formula unit  \\
\hline
0	& 0		& 0.03	& 0.05 	& 0.11 &	Mn$_{0.95}$Bi$_{2.05}$Te$_4$ \\
0.25	& 0.24	& 0.06	& 0.07	& 0.18 &	Mn$_{0.94}$(Bi$_{0.76}$Sb$_{0.24}$)$_{2.06}$Te$_4$    \\
0.5	& 0.5 	& 0.08	& 0.08	& 0.24 &	Mn$_{0.92}$(Bi$_{0.5}$Sb$_{0.5}$)$_{2.08}$Te$_4$ \\
0.75	& 0.75 	& 0.13	& 0.03	& 0.29 &   Mn$_{0.96}$(Bi$_{0.25}$Sb$_{0.75}$)$_{2.04}$Te$_4$   \\
1	& 1 		& 0.16 	& 0.06 	& 0.37 &	Mn$_{0.94}$Sb$_{2.06}$Te$_4$    \\
\hline\hline
\end{tabular}
\label{Rietveld}
\end{table}

\begin{table}
\caption {Structural parameters of Mn(Bi$_{1-x}$Sb$_{x}$)$_2$Te$_4$ determined from Rietveld refinement of POWGEN data at 100 K along with R-factor.  Reported errors of the structural parameters are smaller than the signficant digits of the values shown in the Table.  Values for MBT ($x=0$) are obtained from Ref.~\cite{Yan19}.}
\renewcommand\arraystretch{1.25}
\centering
\begin{tabular}{ c | c | c | c | c | c | c}
\hline\hline
~ $x_{\rm nom}$ ~ & ~$a$~ & ~ $c$ ~ & ~$z_{Te_{1}}$ ~ & ~$z_{Te_{2}}$~ & ~$z_{X}$~ & R-factor \\
\hline
0	& 4.31	& 40.74 	& 0.133	& 0.294	& 0.425	& - 	\\
0.25	& 4.29	& 40.74	& 0.133	& 0.294	& 0.425    	& 7.02 \\
0.5	& 4.27 	& 40.72	& 0.132	& 0.293 	& 0.425 	& 7.74\\
0.75	& 4.24 	& 40.69	& 0.132	& 0.292 	& 0.424   	& 6.99 \\
1	& 4.23 	& 40.65 	& 0.131	& 0.292 	& 0.425    	& 8.97 \\
\hline\hline
\end{tabular}
\label{Lattice}
\end{table}

It has been clearly established that MBT adopts an A-type AF structure where a single Mn layer within a septuple block is FM with moments pointing perpendicular to the layer, but the moment direction is staggered from block-to-block \cite{Yan19}.  Due to the close-packed stacking of Mn layers that results in a doubling of the chemical cell along $c$, this A-type order is characterized by a propagation vector of $\boldsymbol{\tau}=$ (0,0,3/2). 

\begin{figure}
\includegraphics[width=1.0\linewidth]{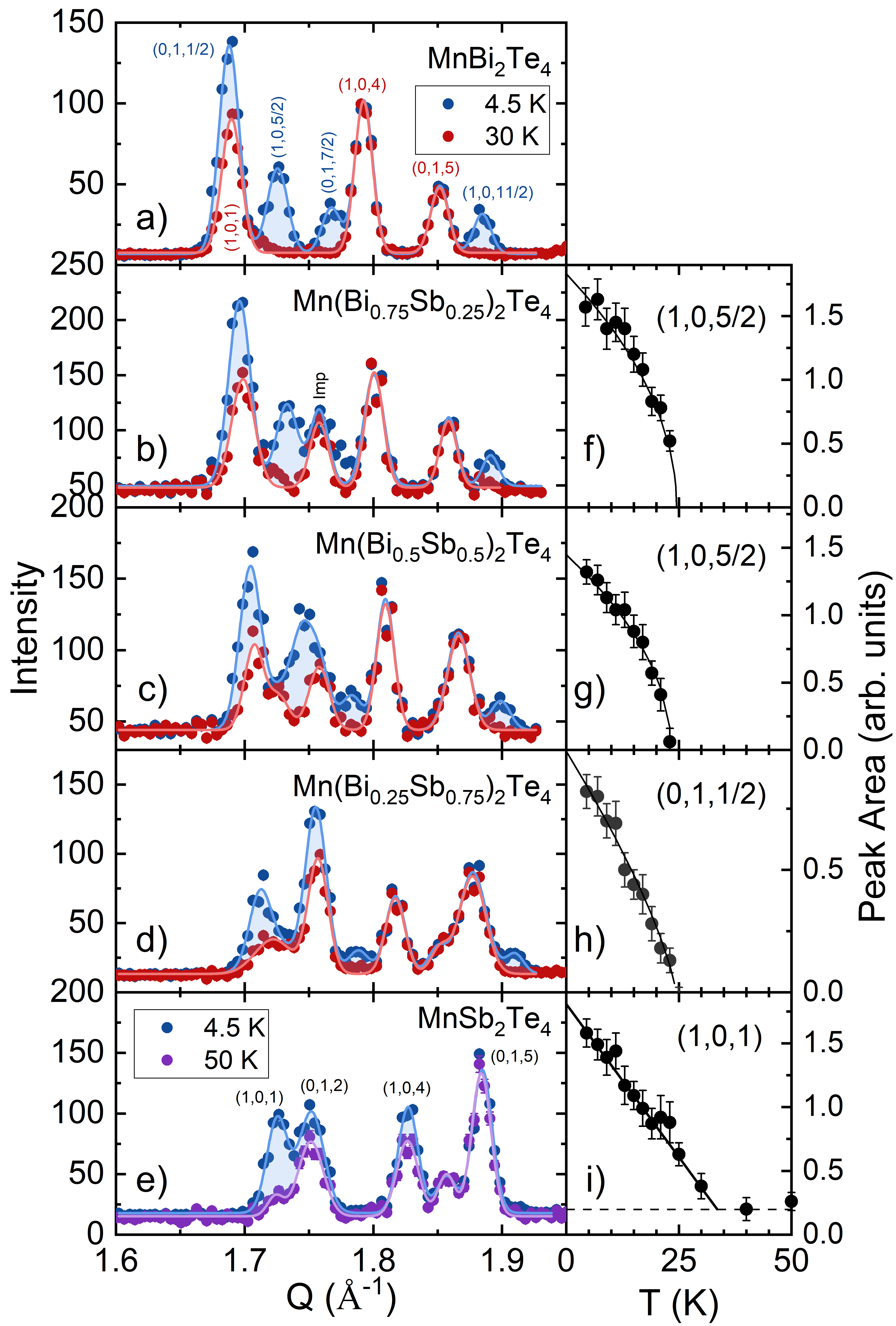}
\caption{\footnotesize (a)-(e) Data from the elastic line of DCS as a function of composition showing diffraction data for nominal compositions $x=$ 0, 0.25, 0.5, 0.75, and 1, respectively.  For (a)-(d), diffraction data is shown above $T_{\rm N}$ ($T=30$ K, red points) and below $T_{\rm N}$ ($T=4.5$ K, blue points). For panel (f) with FM order, data is shown above $T_{\rm C}$ ($T=50$ K, purple points) and below $T_{\rm C}$ ($T=4.5$ K, blue points).  Lines are Gaussian fits to the experimental data and shaded blue area indicates magnetic intensity.  Panels (f)-(i) show the magnetic order parameters for  $x=$ 0.25, 0.5, 0.75, and 1, respectively, as the integrated area of the indicated magnetic peaks.}
\label{fig2}
\end{figure} 

Fig.~\ref{fig2} (a)-(e) shows elastic data of all studied compositions from DCS with $E_i=$ 3.55 meV in the range of $Q=$ 1.6-1.95 \AA~and averaged over an energy window of $\pm$ 0.15 meV.  In this $Q$-range, the main nuclear diffraction peaks are (1,0,1), (0,1,2), (1,0,4), and (0,1,5).  With Sb substitution these peaks shift due to lattice parameter changes  and their intensity varies due to the small shifts in the atomic positions (see Table \ref{Lattice}) and scattering length of Bi and Sb. For example, the (0,1,2) peak is very weak for MBT whereas (1,0,1) is very weak for MST.  

For the A-type structure with moments perpendicular to the layer, magnetic Bragg peaks in this $Q$-range are observed at (1,0,$L \pm 3/2$) and (0,1,$L \pm 3/2$), as shown in Fig.~1(a).   For Sb substituted samples up to $x_{\rm nom}=$~0.75 and measured below $T_{\rm N}$, the A-type structure and its associated propagation vector of (0,0,3/2) is adopted.  However, for $x_{\rm nom}=$1, magnetic intensities are not observed at these positions.  Instead, magnetic intensity characteristic of FM order appears on top of the nuclear Bragg peaks at (1,0,1), (1,0,2) and (1,0,4), as shown in Fig.~1(e).

Fig.~2(f)-(i) show the integrated area of selected magnetic peaks (after subtracting the nuclear scattering) representative of the magnetic order parameter for each composition as a function of temperature.  Data for $x_{\rm nom}=$~0 can be found in Ref.~\cite{Yan19}. The order parameters indicate that the N\'eel temperature up to $x_{\rm nom}=$~0.75 remains approximately constant in agreement with single-crystal and magnetization data \cite{Yan19_2}.  For the FM MST sample, the Curie temperature is significantly higher at $T_{\rm C}=$ 34 K which suggests that the chemical configuration is a significant factor in determining the strength of the magnetic interactions.
  
\section{Analysis of inelastic data}
\subsection{Excitations in magnetically ordered state}
\begin{figure*}
\includegraphics[width=1.0\linewidth]{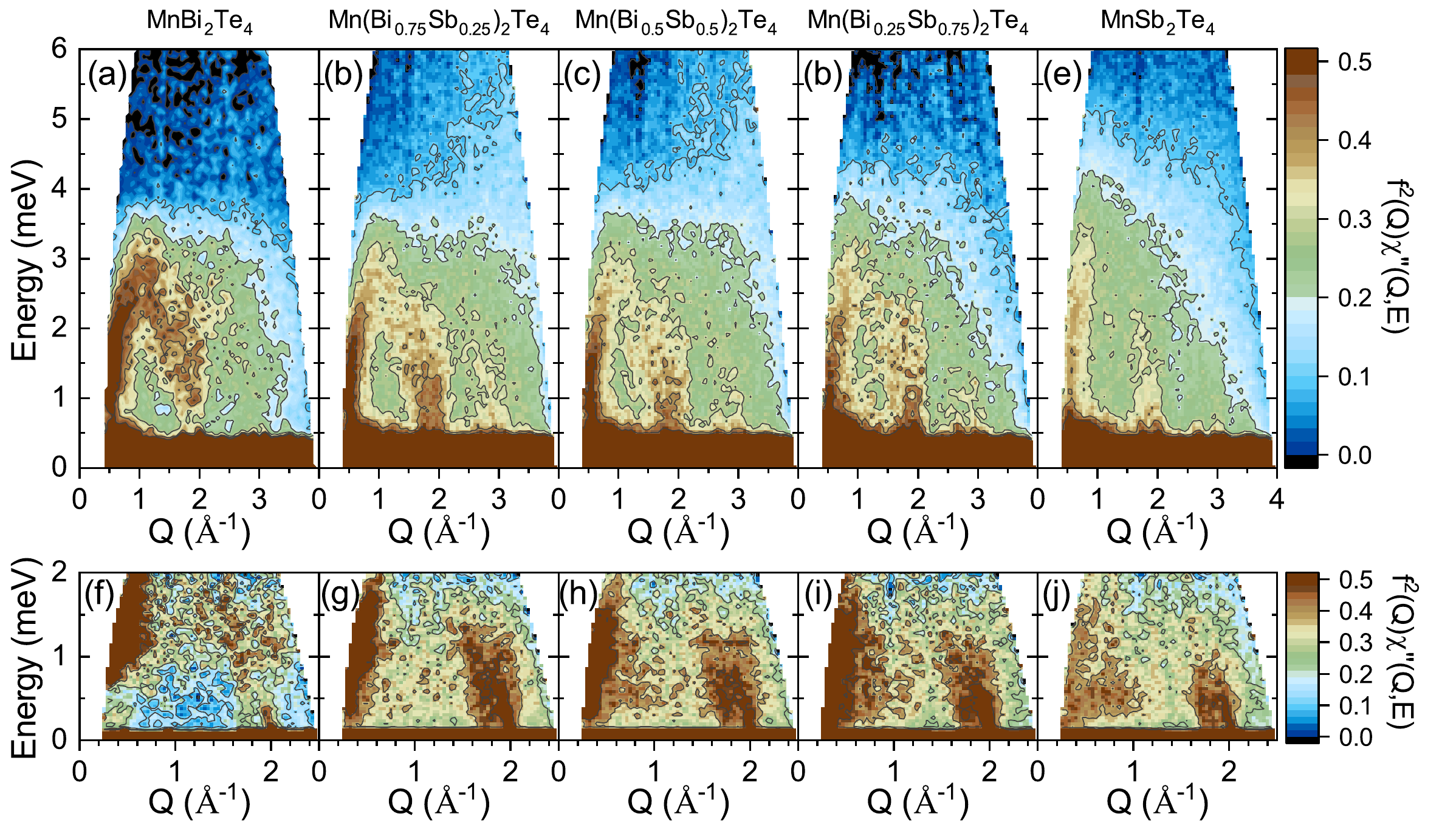}
\caption{\footnotesize Inelastic neutron scattering data from the Mn(Bi$_{1-x}$Sb$_{x}$)$_{2}$Te$_{4}$ series at $T=4$ K.  Panels (a)-(e) show data from $x_{\rm nom}=$ 0, 0.25, 0.5, 0.75 and 1.0 samples, respectively, with $E_{i}=$ 9.09 meV.  Panels (f)-(j) show the same compositional series measured with $E_{i}=$ 3.55 meV.  Color bar indicates intensities reported as the magnetic form factor times the dynamical susceptibility.}
\label{fig_sqe}
\end{figure*}

Figure \ref{fig_sqe} gives an overview of the evolution of the magnetic spectra as a function of $Q$ and $E$ with Sb substitution.  The MnBi$_2$Te$_4$ sample was previously measured using a high intensity instrument and the spectra have been analyzed in detail using a Heisenberg model, as reported previously \cite{Li20}.  The magnetic spectra for MBT measured on DCS at $T=$ 4 K are shown here in Figs. \ref{fig_sqe}(a) and (f) at $E_{\rm i}=$ 9.09 and 3.55 meV, respectively.  These data for MBT are consistent with previous work and consist of a spin gap ($\Delta$) of approximately 0.6 meV with dispersive spin waves propagating to a maximum energy of approximately 3.5 meV. Unlike the work in Ref.~\cite{Li20}, we are not able to clearly observe gap edge oscillations that allow for the determination of weak coupling between Mn layers in adjacent septuple blocks. The spin wave branch bends over and returns to the gap energy at first Brillouin zone center at $Q(1,0,0) \approx 1.7$ \AA$^{-1}$. As described below, Fig.~\ref{fig_models} shows that Heisenberg model calculations using parameters from Ref.~\cite{Li20} agree with current measurements for MBT.    

Figure \ref{fig_sqe} demonstrates two clear features that emerge in the magnetic spectra upon the gradual substitution of Bi with Sb.  First is that the spin gap closes, as expected from the reduction of the critical field for the spin-flop transition \cite{Yan19_2}.  The closure of the spin gap is more evident in the higher-resolution data in Figs. \ref{fig_sqe}(f)-(j).  The second feature is a general broadening of spectral features in both $Q$ and $E$, along with a gradual shift to higher energies.

\begin{figure}
\includegraphics[width=1.0\linewidth]{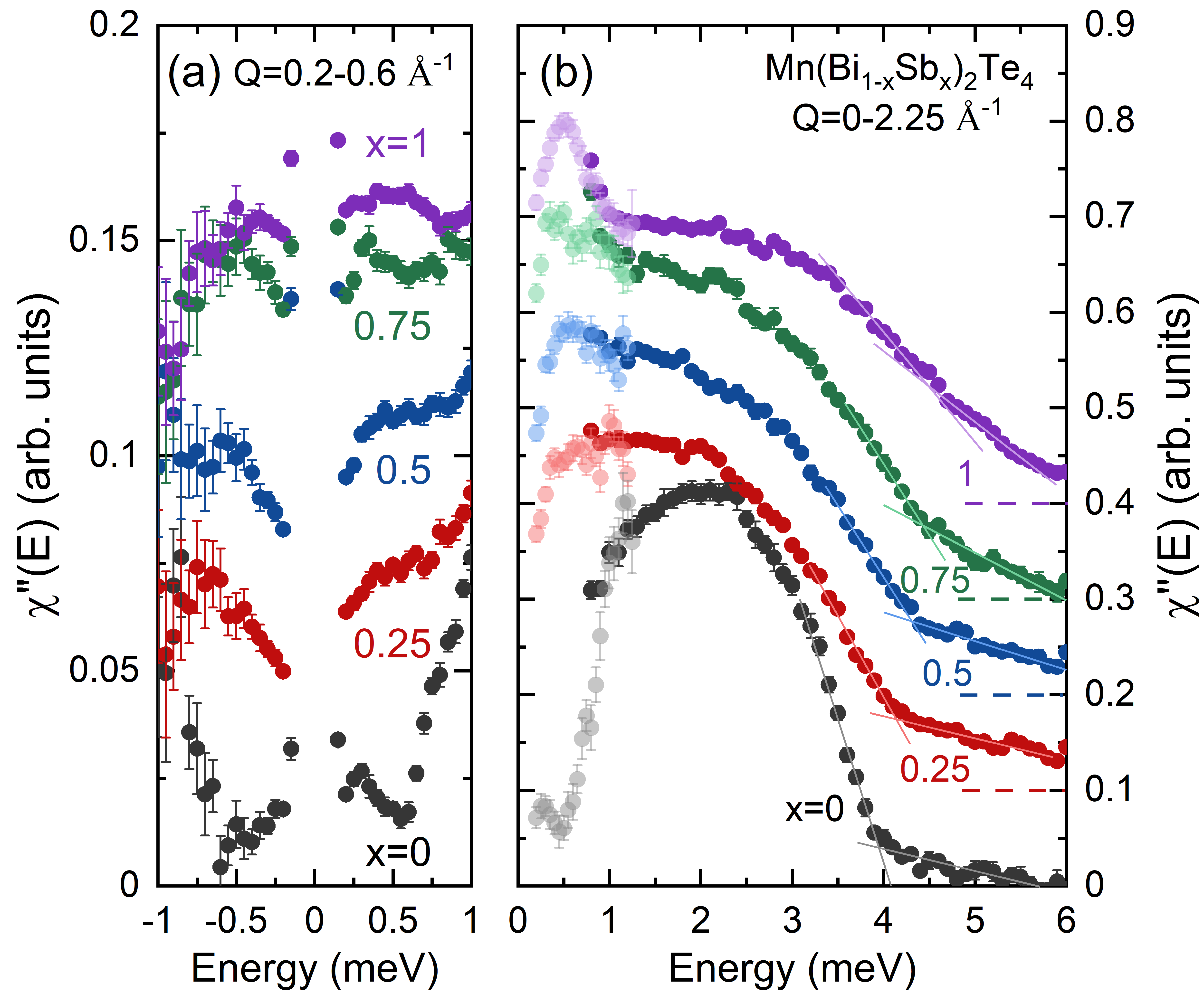}
\caption{\footnotesize  (a) Low-energy cuts of the $E_{\rm i}$=3.55 meV data summed from $Q=$ 0.2--0.6 \AA$^{-1}$ for each composition of Mn(Bi$_{1-x}$Sb$_{x}$)$_{2}$Te$_{4}$ at $T=$ 4 K. (b) The full spectrum of magnetic excitations $E_{\rm i}$=9.09 meV (dark colored symbols) and $E_{\rm i}$=3.55 meV (light colored symbols) summed from $Q=$ 0--2.25 \AA$^{-1}$. In (a) and (b), plots are vertically offset for clarity with baselines shown as short dashed lines in (b).  For each composition in (b), slope line constructions are shown that provide an estimate of the average FM exchange $\langle J_{FM} \rangle$, as described in the text.}
\label{fig_spectra}
\end{figure}

These spectral features are more evident by examining the energy spectra summed over different ranges of $Q$. Figure \ref{fig_spectra}(a) shows low-$Q$ energy cuts over a range from 0.2 to 0.6 \AA$^{-1}$ with $E_{\rm i}=$ 3.55 meV.  Here the spin gap of 0.6 meV in MBT is evident.  Moving to $x=0.25$, already the gap is hard to resolve and likely comparable to, or less than, the elastic energy resolution FWHM of 0.1 meV.  Similar low energy spectral features are found for $x=$ 0.5 and 0.75 compositions.  Moving to MST, the weak gap feature is replaced by the appearance of a significant spectral resonant peak near 0.5 meV that is discussed on more detail below.  

Figure \ref{fig_spectra}(b) shows the full spectra summed from $Q=$ 0 to 2.25 \AA$^{-1}$ for both incident energies.  Low energy features similar to Fig.~\ref{fig_spectra}(a) are seen and connect with a broad magnetic spectrum extending above 4 meV.  For MBT, a fairly sharp cutoff occurs near 4 meV followed by a weak tail due to Lorentzian damping.  A numerical estimate of the upper bandwidth cutoff for the spin waves can be obtained by finding the energy ($E_{\rm max}$) where the spectra change slope in the energy range from 3--5 meV.  With increasing $x$, $E_{\rm max}$ shifts to higher energy, but the feature broadens out significantly suggesting a concomitant growth of the damping. For MST, the upper cutoff feature is obscured and replaced by a long tail that extends up to approximately 6 meV.  The general trend of a magnetic energy scale that increases with Sb substitution can also be captured by the determination of the first and second energy moments of the spectra, given by
\begin{equation}
\langle E^n \rangle = \frac{\sum_Q \sum_E E^n S(Q,E)}{\sum_Q \sum_E}.
\end{equation}
All of the estimated magnetic energy scales are listed in Table \ref{energy_table} and demonstrate a general increase in the energy scale of the magnetic excitations with the replacement of Bi by Sb.
\begin{table}
\caption {Spin gap, ($\Delta$), the top of the energy band $E_{\rm max}$ and the first and second energy moments of the neutron scattering intensities of Mn(Bi$_{1-x}$Sb$_x$)$_2$Te$_4$ obtained from $E_{\rm i}=$ 9.09 meV data and averaged over $E=$1--6 meV and $Q=$ 0--2 \AA$^{-1}$.  All energy values appear in meV.}
\renewcommand\arraystretch{1.25}
\centering
\begin{tabular}{ c | cccc }
\hline\hline
~ $x_{nom}$ ~  & ~ $\Delta$ ~ & ~$E_{\rm max}$ ~  &  ~$\langle E \rangle$ ~ &  ~$\sqrt{\langle E^2 \rangle}$~  \\
\hline
0	& 0.6 		& 4.0 	& 2.27 (3) & 2.42 (2)   \\
0.25	& $\sim0.1$ 	& 4.1 	& 2.51 (2) & 2.78 (2)  \\
0.5	& $\sim0.1$	& 4.3 	& 2.54 (2) & 2.75 (2) \\
0.75	& $\sim0.1$ 	& 4.4 	& 2.53 (2) & 2.75 (2)   \\
1	& $\sim0.$ 	& 4.5 	& 2.81 (2) & 3.06 (1) \\
\hline\hline
\end{tabular}
\label{energy_table}
\end{table}

\subsection{Modeling of magnetic spectra of MnBi\texorpdfstring{\textsubscript{2}}\ Te\texorpdfstring{\textsubscript{4}}~~and doped compositions}

We now turn to Heisenberg modeling of the data. Given the broad spectral features, especially considering the heavy chemical disorder for intermediate Sb concentrations, the detailed development of such models is impractical.  Thus, here we focus only on a comparison of the general features of Heisenberg models for MBT and MST.  

\begin{figure}
\includegraphics[width=1.\linewidth]{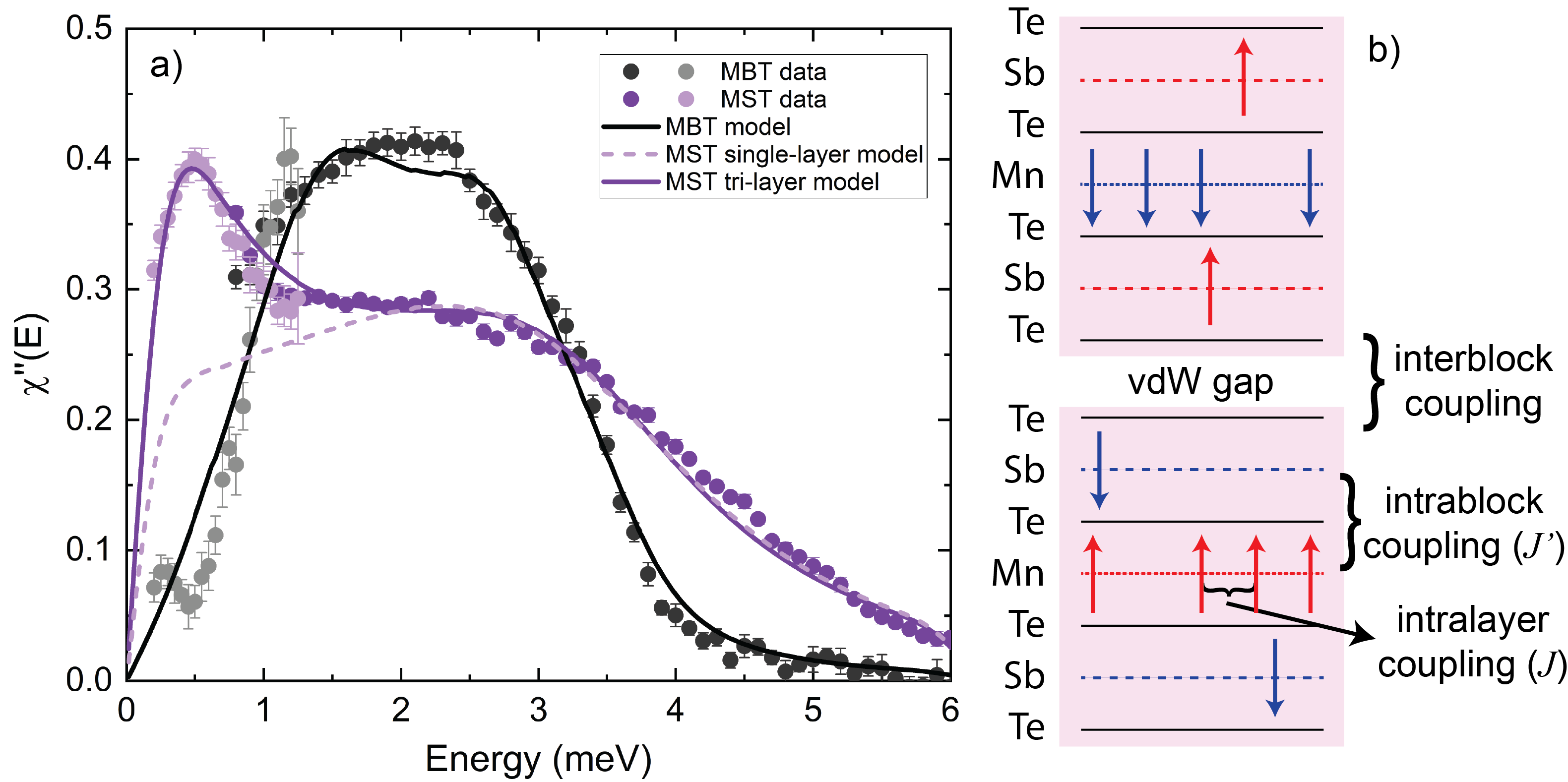}
\caption{\footnotesize  a) Comparison of Heisenberg model fits of MnBi$_2$Te$_4$ and MnSb$_2$Te$_4$ (lines) to the magnetic spectra obtained from DCS (symbols).  For MBT (black solid line), the Heisenberg parameters are obtained from Ref.~\cite{Li20}.  For MST, there are two model curves.  The purple dashed line is a single-layer model using MBT intralayer interactions with all FM interactions increased in strength by 50\% and a large uniform damping of $\Gamma=3$ meV (dashed purple line).  The solid purple line is a tri-layer ferrimagnetic model with parameters described in the text. b) Schematic diagram of the ferrimagnetic structure of MST. Individual septuple blocks are shaded in pink and separated by a van der Waals (vdW) gap. Intralayer FM coupling ($J$), intrablock AF coupling ($J'$) and AF interblock coupling are indicated.}
\label{fig_models}
\end{figure}

For MBT, we use the published parameters from Ref.~\cite{Li20} and compare them to the energy spectrum measured on DCS in Fig.~5.  This model contains a single interlayer coupling ($SJ_c=-0.055$ meV), intralayer couplings up to the fourth-nearest neighbor ($SJ_1=0.3$, $SJ_2=-0.08$, and $SJ_4=0.023$ meV), and a uniaxial anisotropy parameter ($SD=0.12$ meV).  In addition, this model also includes significant intrinsic spin wave broadening in the form of a damped simple harmonic oscillator (DSHO) lineshape with $\Gamma=0.7$ meV.  The damping appears to become even larger in Sb substituted samples.  In Fig.~\ref{fig_models}, we calculate the polycrystalline-averaged MBT Heisenberg spectrum using these parameters \cite{McQueeney08} and the appropriate resolution and instrument configuration for DCS at $E_{\rm i}=9.09$ meV and compare to the experimental data.  The agreement is reasonable, as expected.  

The development of a similarly detailed Heisenberg model for Sb-substituted compositions is not possible due to increased magnetic disorder and severe broadening.  However, we can discuss simplified models using the energy scale $E_{\rm max}$ from Table \ref{energy_table}.  Studies of numerical spin wave models show that $E_{\rm max}$ can be used to estimate the average intralayer FM coupling, $E_{\rm max} \approx \Delta + 9S\langle J_{FM} \rangle$.  Previous INS data find such couplings to have long-range character \cite{Li20, Li20_2}, requiring the average FM coupling parameter to be represented by a sum over coordination shells $\langle J_{FM} \rangle = \frac{1}{6} \sum_i z_i J_i^{FM}$.    Antiferromagnetic intralayer couplings that are present do not affect $E_{\rm max}$, but rather redistribute the magnetic spectral weight below $E_{\rm max}$.  

Using this rough estimate and the parameters from Ref.~\cite{Li20}, we obtain $E_{\rm max}({\rm MBT}) = 3.7$ meV, which is close to our experimental determination in Table \ref{energy_table}. Conversely, we can use the measured $E_{\rm max}$ to estimate $S\langle J_{FM} \rangle ({\rm MBT}) =$ 0.37 meV. Based on this analysis, we may be tempted to assume that the $\langle J_{FM} \rangle$ is increasing with Sb substitution.  This general trend is supported by the visible change in the stiffness of the principal magnon branch, as shown in a comparison of Figs.~\ref{fig_sqe}(a) and (e), as well as first-principles calculations described below.  However, appreciable antisite mixing with heavy Sb substitution introduces a large AF coupling between layers within the septuple block (intrablock coupling) that results in partially compensated ferrimagnetism, as shown in Fig.~\ref{fig_models}.  This special ferrimagnetic case is applied to MST below.

\section{Ferrimagnetism in M\texorpdfstring{\lowercase{n}S\lowercase{b}\textsubscript{2}T\lowercase{e}\textsubscript{4}}{TEXT}}
The simplest approach to develop a Heisenberg model for MST is to set $SD=SJ_c=0$ to account for the vanishing spin gap and the lack of features derived from the interlayer dispersion, as described in Ref.~\cite{Li20}.  In this case, the model is that of a 2D triangular FM sheet which makes no distinction between 3D FM (MST) and A-type AF (MBT) states.  We estimate the intralayer FM exchange values for MST by rescaling the MBT values upwards by 35\% based on the $S\langle J_{FM} \rangle$ expression.  We leave $SJ_2$ unchanged from its MBT value.  Using these Heisenberg parameters, the agreement of the model and data for MST is very poor.  Increasing the average FM coupling to be 50\% larger than MBT and significantly increasing the damping to a value of $\Gamma=3$ meV provide much better agreement of the high-energy tail above 2 meV. This single-layer FM model for MST is compared to the data in Fig.~\ref{fig_models}.

However, what remains unresolved by the single-layer model is the low energy resonance near 0.5 meV, as shown in Figs.~\ref{fig_spectra} and \ref{fig_models}.  A possible origin of this peak is ferrimagnetism within a single septuple block that originates from Mn/Sb antisite mixing \cite{Murakami19}.  In the ferrimagnetic model for MST, each septuple block consists of three magnetic layers, the principal Mn layer (with associated Sb$_{\rm Mn}$ magnetic vacancies) and two Sb layers with small concentrations of Mn$_{\rm Sb}$ moments.  Antiferromagnetic coupling between ions in the main Mn layer and the Mn$_{\rm Sb}$ ions in the Sb layers (AF intrablock coupling) will generate a ferrimagnetic septuple block with reduced magnetization, as shown in Fig \ref{fig_models}. Similar to the spin dynamics of other ferrimagnets, out-of-phase spin precessions of the three layers can result in localized and dispersionless (optical) spin excitations which may appear as a resonant-like feature in the magnetic spectra.


We develop a simple model to account for tri-layer ferrimagnetism in MST based on the schematic structure shown in Fig \ref{fig_models}. The model consists of additional close-packed layers ($A$ and $C$) above and below the main FM Mn layer ($B$) which have opposite magnetization.  In this model, we do not include real effects of magnetic vacancy and impurity disorder.  Instead, we assume that each Mn in the main $B$ layer (layer $B$) has an average spin $S_B=S$ and an intralayer FM coupling $J>0$. In the $A$ and $C$ layers, we assume that the average spin is reduced $S_A=S_C=s<S/2$ to account for the compensated magnetization, $M \propto S-2s$,  caused by AF intrablock exchange that couples $AB$ and $BC$ layers ($J'<0$). We assume that there is no intralayer coupling in the $A$ and $C$ layers.  Within linear spin wave theory of the tri-layer ferrimagnetic Heisenberg model, there are three branches with dispersion
\begin{multline}
\omega_1({\bf q})=\frac{1}{2}[3J'(S-2s) +6SJ\gamma({\bf q})] \\ +\frac{1}{2}\sqrt{[3J'(S-2s)+6SJ\gamma({\bf q})]^2-24 S J' (3SJ-2sJ' )\gamma({\bf q})}
\end{multline}
\begin{multline}
\omega_2({\bf q})=\frac{1}{2}[3J'(S-2s)+6SJ\gamma({\bf q})] \\ -\frac{1}{2}\sqrt{[3J'(S-2s)+6SJ\gamma({\bf q})]^2-24 S J' (3SJ-2sJ' )\gamma({\bf q})}
\end{multline}
\begin{equation}
\omega_3({\bf q})=3S|J'|.
\end{equation}
We define the function 
\begin{equation}
\gamma({\bf q})=1-\frac{1}{3}\big[\mathrm{cos}(\textbf{q} \cdot \textbf{a}_{1})+\mathrm{cos}(\textbf{q} \cdot \textbf{a}_{2})+\mathrm{cos}(\textbf{q} \cdot (\textbf{a}_{1}+\textbf{a}_{2}))\big]
\end{equation}
where $\textbf{a}_{1}=a\hat{x}$ and $\textbf{a}_{2}=-\frac{1}{2}a\hat{x}+\frac{\sqrt{3}}{2}a\hat{y}$ define the triangular layer unit cell.

In the trilayer model, $\omega_1({\bf q})$ is the acoustic branch [$\omega_1(0)= 0$] with a bandwidth of $9SJ+6s|J'|$ that reduces to the FM single-layer dispersion when $J'=0$.  $\omega_2({\bf q})$ is an optical branch with zone center energy of $\omega_2(0)= 3|J'|(S-2s)$ and bandwidth of $6s|J'|$ (i.e. it is nearly dispersionless for dilute $A$ and $C$ layers when $s$ is small). The $\omega_3$ mode is completely dispersionless and has $+0-$ symmetry, corresponding to oppositely precessing $A$ and $C$ layers with a silent $B$ layer.  Thus, $\omega_2$ and $\omega_3$ form a nearly dispersionless excitation band which we associate with the low-energy resonance mode.

Using the simple relations above, we can estimate an initial set of parameters for the trilayer ferrimagnetic model.  For an antisite mixing concentration $y$ with $M \propto S-2s$,  we estimate that $s/S \approx y/(1-2y) \approx$ 0.25. Numerical studies find the localized resonance peak at $\omega_2$ giving $S|J'| \approx$ 0.35 meV. Finally, since the overall bandwidth of spin excitations is increased by $J'$, we modify the relation for the maximum energy scale; $E_{\rm max} \approx  9SJ+6s|J'|$ which results in an estimate of $SJ \approx$ 0.45 meV. 

After making small adjustments to these parameters, we obtain reasonably good agreement with the data for $s/S=0.23$,  $SJ'=-0.3$ meV and $SJ=0.37$ meV, as shown in Fig.~\ref{fig_models}.  We cannot take the tri-layer model parameters to represent the actual pairwise exchange values as the magnetic spectra are affected by different magnetic configurations due to significant magnetic disorder.  This disorder is also likely reflected in the substantial DSHO width of $\Gamma=$ 3 meV. However, based on these simple assumptions, it is reasonable to assign the low-energy resonance peak to a ferrimagnetic localized mode originating from strong AF coupling of the main Mn layer to antisite impurity Mn ions in the Sb layer.

\section{Magnetic excitations in paramagnetic state}
\begin{figure}
\includegraphics[width=0.7\linewidth]{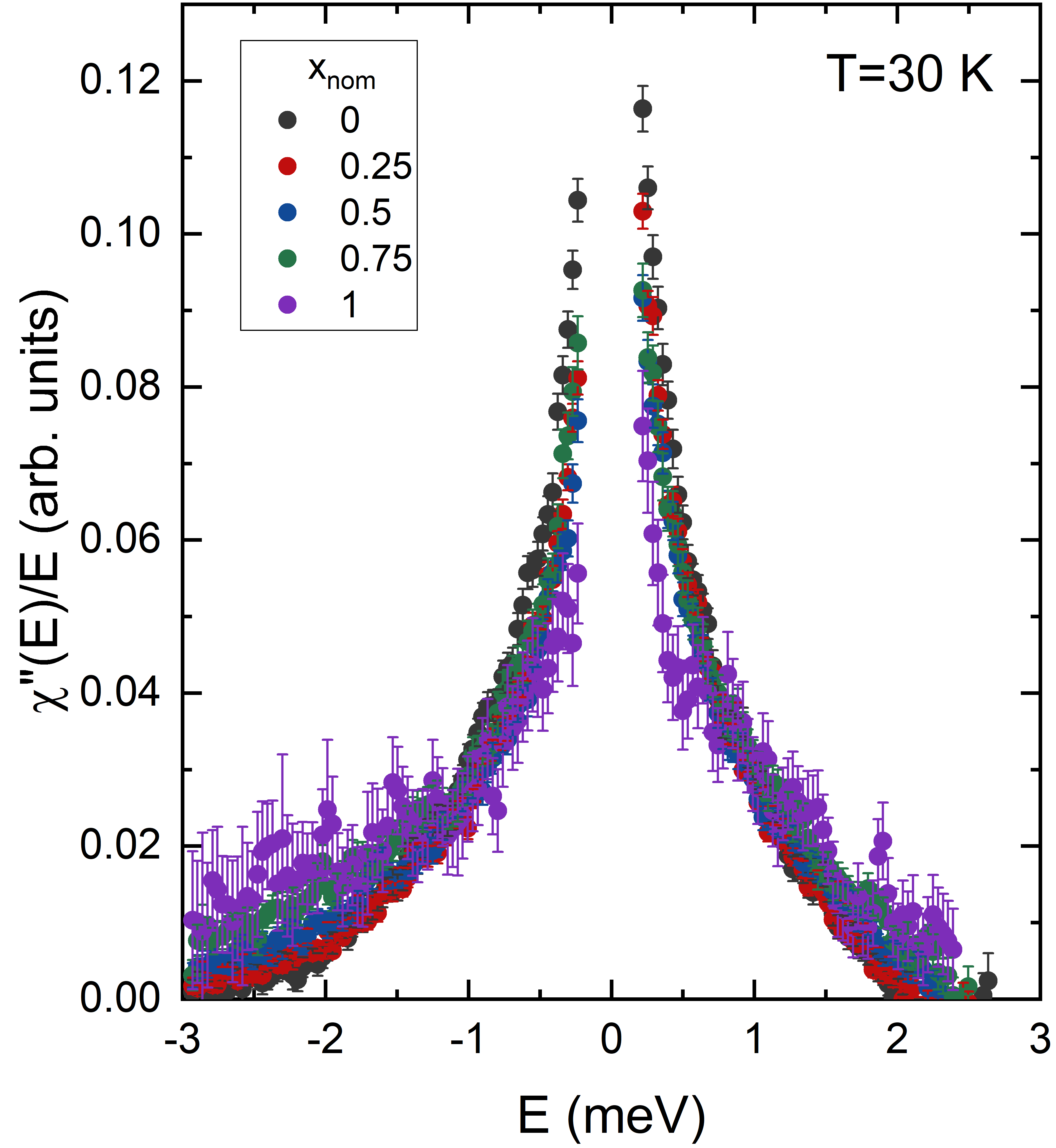}
\caption{\footnotesize  Low-energy cuts of the $E_{\rm i}$=3.55 meV data summed from $Q=$ 0--1.25 \AA$^{-1}$ for each composition of Mn(Bi$_{1-x}$Sb$_{x}$)$_{2}$Te$_{4}$ at $T=$ 30 K.}
\label{fig_30K}
\end{figure}

Figure \ref{fig_30K} shows the magnetic spectra plotted as $\chi"(E)/E$ at $T=$ 30 K for all compositions. Typically, magnetic spectra evolve into a relaxational Lorentzian lineshape above the magnetic ordering temperature where the Lorentzian half-width-at-half-maximum ($\gamma$) represents the spin relaxation rate.  For the $x=0$--0.75 compositions, reasonable fits to a Lorentzian lineshape are obtained and have identical Lorentizan half-width-at-half-maximum of $\gamma=$0.67(2) meV.  Unconstrained fits to the MST spectrum find a different width of $\gamma=$ 0.85(6) meV, or about 25\% larger. However, the MST spectra is qualitatively different and the localized resonance mode remains near 0.5 meV.  This suggests that AF correlations between Mn and Mn/Sb sites persists above the ordering temperature.  

\section{First-principles calculations of the magnetic exchange interactions}
We carry out DFT+$U$~\cite{Dudarev98} calculations to estimate the effective exchange and anisotropy parameters of the considered model spin Hamiltonian.
Calculations are performed within the generalized gradient approximation using the exchange-correlation functional of Perdew, Burke, and Ernzerhof~\cite{PBE96} as implemented in \textsc{vasp}~\cite{Kresse96, Kresse99}.
The nuclei and core electrons were described by the projector augmented wave potential~\cite{Blochl94}.
Plain DFT calculations predict wrong the AF intralayer ordering in MBT and additional electron-electron repulsion on Mn-$d$ orbitals is needed to describe the magnetic interactions properly in these systems.
Here, we apply $U$=3--5~eV, which is shown to predict the correct $A$-type AFM ground state in MBT~\cite{Li20}.
The plane wave cutoff energy is set at 346 eV, and the self-consistent-field energy threshold is set to 10$^{-6}$ eV.

To estimate the role of the single-ion anisotropy in the spin gap, we calculate the magnetocrystalline anisotropy energy (MAE) $K= E_{a}{-}E_{c}$ in the AF ground state of both MBT and MST, where $E_{a}$ and $E_{c}$ are the total energies of the system with spins aligned along the $a$ or $c$ axis, respectively.
SOC is included using the second-variation method~\cite{li1990prb}.

Table \ref{dft} shows no change in calculated Mn moments in the A-type AF state for stoichiometric MBT and MST.
However, DFT finds a significant lowering of the MAE by more than a factor of two when comparing MBT and MST.
This is caused by the lower SOC of Sb relative to Bi and certainly contributes to the observed closure of the spin gap.

Three spin exchanges, $J_1$, $J_2$, and $J_c$ were investigated for stoichiometric MBT and MST by mapping the total energy of various collinear spin configurations into  the Heisenberg spin Hamiltonian
\begin{equation}
H= -\sum_{\langle ij \rangle}J_{ij}  \textbf{S}_{i} \cdot \textbf{S}_{j}, 
\label{heisenberg}
\end{equation}
where $J_{ij}$ are the spin exchanges $J_1$, $J_2$, and $J_c$, and $S$ represents the total spin on Mn$^{2+}$ (i.e., $S=$ 5/2).
A supercell of 12 formula units (f.u.) is used to accommodate four collinear spin configurations.


Table \ref{dft} shows the exchange values determined for both MST and MBT.
DFT predicts a 23-33\% increase in the FM nearest-neighbor interaction  $J_1$ for MST.
On the other hand, the competing AF next-nearest-neighbor interaction $J_2$ is 10-30\% lower for MST, which somewhat lessens the effect of magnetic frustration.
DFT predicts a $\sim 20$\% larger AF interlayer coupling $J_c$ in MST.

\begin{table}[htb]
  \caption{On-site magnetic moment $m$, magnetocrystalline anisotropy energy $K$, and exchange couplings $J_1$, $J_2$,  and $J_c$ in stoichiometric MnBi$_2$Te$_4$ and MnSb$_2$Te$_4$.
    DFT+$U$ calculations were performed with $U$=3 and 5~eV.}
\label{dft}
\bgroup
\def\arraystretch{1.3} 
\begin{tabular*}{\linewidth}{c @{\extracolsep{\fill}}ccccccccc}
\hline\hline
Compound          & & $U$   &  $m$                & $K$          & $J_1$ & $J_2$ & $J_c$   \\   \cline{6-8}
                  & & (eV)  &  ($\mu_{\rm B}$/Mn) &  (meV/f.u.)  &         & (meV) &     \\ \hline
MnBi$_2$Te$_4$    & & 3     &4.48 & 0.444 & 0.220 & -0.054 & -0.035   \\
MnBi$_2$Te$_4$    & & 5     &4.59 & 0.329 & 0.281 & -0.021 & -0.010   \\ \hline
MnSb$_2$Te$_4$    & & 3     &4.48 & 0.202 & 0.292 & -0.050 & -0.042    \\
MnSb$_2$Te$_4$    & & 5     &4.59 & 0.151 & 0.317 & -0.016 & -0.012    \\
\hline\hline
\end{tabular*}
\egroup
\end{table}

While some of the trends observed in the DFT calculations for MBT and MST correspond qualitatively to observations (and some do not), we caution that the presence of substantial magnetic disorder in MST must be considered.
The most obvious omission in stoichiometric DFT calculations is the antisite-defect-driven ferrimagnetism and its overall effect on the magnetic energy scales.
For example, the increased FM $J_1$ obtained from DFT for MST is consistent with the 35\% increase of $\langle J_{\rm FM} \rangle$ estimated from the experimental data.
However, we also know that these average quantities can also be affected by ferrimagnetic coupling.
Also, DFT predicts MST AF $J_c$, whereas sample growths can generate ground states with either AF or FM interblock ordering.
This signifies a strong sensitivity of $J_c$ to magnetic vacancies and disorder, and it is possible that Mn$_{\rm Sb}$ ions have a significant effect on the coupling across the van der Waals gap.

\begin{figure}
\includegraphics[width=1\linewidth]{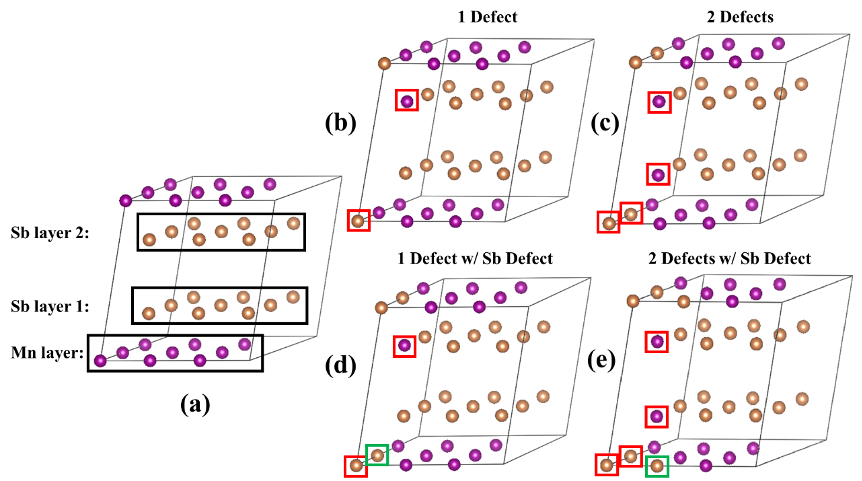}
\caption{\footnotesize Supercell (a) of MnSb$_2$Te$_4$ used to create structures with Mn$_{\rm Sb}$ antisite defects.
Mn (purple) and Sb (orange) atoms shown.
Structures correspond to (b) one antisite defect, (c) one antisite defect with one additional Sb$_{\rm Mn}$ magnetic vacancy, (d) two antisite defects, and (e) two antisite defects with one additional Sb$_{\rm Mn}$ magnetic vacancy.
Red boxes correspond to atomic positions where antisite defect are placed, and green boxes correspond to atomic positions of Sb$_{\rm Mn}$.}
\label{DFT_defects}
\end{figure}

To study the effect of disorder on the magnetic interactions in MST, we employ a supercell of $3a \times 3b \times 1c$ formula units 
(one septuple block) with different magnetic defect structures.
Four defect structures are considered: one Mn/Sb antisite defect, one Mn/Sb antisite defect with one additional Sb$_{\rm Mn}$ magnetic vacancy, two Mn/Sb antisite defects, and two Mn/Sb antisite defects with one additional Sb$_{\rm Mn}$ magnetic vacancy.
These structures are shown in Fig.~\ref{DFT_defects}.
With experimental lattice constants fixed, the atomic positions were relaxed with a force criterion of 0.02~eV/\AA.
The location of these defects was chosen to minimize the total energy of each structure.
The $U$ value of 5 eV, which was found to describe the magnetic interaction and spin excitations in pristine MBT well~\cite{Li20_2}, was used in all of the structure relaxation and following magnetic energy calculations.

The effects of these defect configurations on the intrablock magnetic coupling were determined by comparing the calculated total energies for a fully FM septuple block (FM intrablock coupling) with a ferrimagnetic block (AF intrablock coupling). Experimental evidence clearly finds ferrimagnetic blocks. 
Table \ref{DFT_defect_table} summarizes the intrablock magnetic energies of defect configurations shown in Fig.~\ref{DFT_defects}. 
Surprisingly, DFT+$U$ finds that a single antisite defect structure has FM intrablock coupling.
Introducing a second antisite defect or an additional Sb$_{\rm Mn}$ defect, especially the latter, causes the intrablock coupling to become less FM.
With two antisite defects and one Sb$_{\rm Mn}$ defect in the supercell, the intrablock coupling becomes AF.
These results clearly suggest the importance of Sb$_{\rm Mn}$ magnetic vacancies and the binding of antisite defects to promote the intrablock ferrimagnetism that is observed experimentally.  If true, this could suggest that getting rid of Sb$_{\rm Mn}$ vacancies may promote FM intrablock magnetism.  
However, it is worth noting that smaller $U$ values make the intrablock coupling more AF.
\begin{table}[htb]
  \caption{Intrablock magnetic energies $E_{\rm M}$ of various defect configurations in MST. Calculations were carried out in DFT+$U$ using $U =$ 5 eV. $E_{\rm M}$ is calculated as the energy difference between the AF and FM intrablock configurations, $E_{\rm AF} - E_{\rm FM}$. Positive (negative) $E_{\rm M}$ values correspond to FM (AF) intrablock couplings.}
\label{DFT_defect_table}
\bgroup
\def\arraystretch{1.3}                        
\begin{tabular*}{\linewidth}{lr}
\hline\hline
Defect Structure 		        & $E_{\rm M}$ (meV/supercell) \\ \hline
Antisite Defect 	                &	5.57 \\
Antisite Defect + Sb$_{\rm Mn}$		& 	0.97 \\
Two Antisite Defects			&	4.09 \\
Two Antisite Defects + Sb$_{\rm Mn}$	&      -2.22 \\
\hline\hline
\end{tabular*}
\egroup
\end{table}

Finally, the interblock magnetic coupling of the defect structure with two antisite defects and one Sb$_{\rm Mn}$ defect (the defect configuration that clearly promotes ferrimagnetic septuple blocks) is investigated.
To study interblock coupling, the original single septuple block supercell is doubled along the $c$-axis to simulate the four spin configurations with different combinations of the intrablock and interblock orderings.
As shown in Table \ref{DFT_defect_table2}, the lowest energy magnetic structure occurs when the intrablock ordering is AF, and the interblock ordering is FM.
Thus, defects can also stabilize the FM interblock coupling that is observed in some MST samples.
Recall that, for MST, both AF and FM interblock ordering are observed experimentally, which again suggests that microscopic details of the magnetic defect configurations control the global magnetic structure in MST.
Hence, DFT$+U$ calculations confirm that the increasing the defect concentration could stabilize the AF intrablock coupling and FM interblock coupling in MST.

\begin{table}[htb]
  \caption{The energies of the MST defect structure with two antisite defects and one Sb$_{\rm Mn}$ defect. Four different spin configurations are considered.  The energy of the FM-intrablock-and-FM-interblock magnetic structure is chosen as the reference energy zero.}
\label{DFT_defect_table2}
\bgroup
\def\arraystretch{1.3} %
\begin{tabular*}{\linewidth}{cc}
\hline\hline
Defect Structure 					& $E_{\rm M}$ (meV/supercell) \\
\hline
FM intrablock; FM interblock	&	0		\\
FM intrablock; AF interblock	&	4.35		\\
AF intrablock; FM interblock	&	-4.49		\\
AF intrablock; AF interblock	&	-1.57		\\
\hline\hline
\end{tabular*}
\egroup
\end{table}

\section{Discussion}
Taking into account now the overall trends presented here, we start to understand the evolution of magnetism with Sb substitution.  The four main effects are: (1) the introduction of chemical and magnetic disorder, primarily in the form of antisite mixing of Mn and Sb ions, (2) rapid closure of the spin gap and the growth of low energy magnetic spectral weight, (3) an increase in the magnetic energy scale, and (4) an increase in the spectral broadening.  

With respect to (1), it is of course certain that chemical disorder is present based on Bi/Sb alloying.  In addition, our diffraction data confirm that \textit{magnetic} disorder in the form of Mn/(Bi,Sb) antisite exchange is present in all samples and similar to that reported for MST \cite{Murakami19}.  

With respect to (4), both of sources of disorder can be expected to cause significant modifications of the magnetic spectra.  Spectral broadening and extreme damping across the series can have several origins, such as magnon-magnon coupling, Landau damping (in metals), or chemical/magnetic disorder.  For Bi/Sb disorder, DFT calculations in stoichiometric MBT and MST point to site-to-site variation in the intralayer FM coupling strength, although in principle, this source of spectral broadening is absent in MST.  Thus, the development of signficant Mn/Sb antisite disorder and associated Mn layer vacancies are the primary candidate for explaining the growth of spectral broadening.  

With respect to (2), the closure of the gap is expected based on the observed lowering of the critical field for the spin-flop transition.  Part of the growth in the low energy spectral weight arises from the gap closure. However, the sharp resonance peak, most prominent in MST, also likely derives from Mn/Sb antisite disorder resulting in ferrimagnetic septuple block with strong intrablock AF coupling.  

Finally, with respect to (3), it is likely that the increased energy scale across the series arises from a combination of an increase in the intralayer FM coupling and intrablock AF coupling between magnetic defects and the main FM Mn layer. In general, magnetic spectra of MST and Sb-substituted MBT are complex, consisting of heavily damped collective and localized excitations that present many challenges for both experiments and first-principles calculations.

\section{Conclusion}
Overall, these data find that tuning the MBT-MST series through the charge neutral point (near $x \approx 0.3$) while closing the inverted band gap is a promising method to achieve interesting and technologically important magnetic topological states.  However, real materials issues, in the form of chemical and magnetic disorder are widespread.  In addition to having potential consequences on the electron mobility for topological transport, such defects also have non-trivial consequences on the magnetic state itself and the magnetic energy scales.  Variability in finding either the FM or AF magnetic ground state of MST is proof that different chemical and magnetic configurations can affect the global magnetic symmetry.  Also, the tendency for antisite mixing to compensate the FM layers and reduce the net magnetization may have consequences on the exchange coupling to fermions that principally drive topological phase transitions (eg. from AF topological insulator to FM Weyl semimetal).  Finally, the spin excitations may provide a unique window to study chiral charge carriers in FM Weyl semimetals through spin-fermion coupling \cite{Liu13}. The search for such unique responses requires careful analysis of dispersion and lineshape anomalies.  At least for MST, it would seem that chemical and magnetic disorder would make such studies difficult.

\section{Acknowledgments}
RJM would like to thank C. M. Brown for assistance and hospitality with the NIST experiment. We also thank Wei Zhou for his experimental support with the DCS instrument at NIST. Work at Ames Laboratory and Oak Ridge National Laboratory was supported by the U.S. Department of Energy (USDOE), Office of Basic Energy Sciences, Division of Materials Sciences and Engineering. Ames Laboratory is operated for the USDOE by Iowa State University under Contract No. DE-AC02-07CH11358. We acknowledge the support of the National Institute of Standards and Technology, U.S. Department of Commerce, in providing the neutron research facilities used in this work.  A portion of this research used resources at the Spallation Neutron Source, which is a DOE Office of Science User Facility operated by the Oak Ridge National Laboratory.

\end{document}